\documentstyle{elsart}

\input epsf 

\begin{document}

\sloppy

\begin{frontmatter}

\title{On the optimum spacing of stereoscopic imaging atmospheric Cherenkov
telescopes}

\author[1]{W.~Hofmann$^*$},
\author[1]{G.~Hermann},
\author[1]{A.~Konopelko},
\author[1]{H.~Krawczynski},
\author[1]{C.~K\"ohler},
\author[1]{G.~P\"uhlhofer},
\author[1]{F.A.~Aharonian},
\author[2]{A.G.~Akhperjanian},
\author[1]{M.~Aye},
\author[3,4]{J.A.~Barrio},
\author[1,9]{K.~Bernl\"ohr},
\author[4]{J.J.G.~Beteta},
\author[6]{H. Bojahr},
\author[4]{J.L.~Contreras},
\author[4]{J.~Cortina},
\author[1]{A.~Daum},
\author[5]{T.~Deckers},
\author[3,4]{J.~Fernandez},
\author[4]{V.~Fonseca},
\author[4]{J.C.~Gonzalez},
\author[7]{G.~Heinzelmann},
\author[1]{M.~Hemberger},
\author[1]{A.~Heusler},
\author[1]{M.~He\ss},
\author[6]{H.~Hohl},
\author[3]{I.~Holl},
\author[7]{D.~Horns},
\author[1]{I.~Jung},
\author[1,2]{R.~Kankanyan},
\author[3]{M.~Kestel},
\author[1]{J.~Kettler},
\author[1]{A.~Kohnle},
\author[3]{H.~Kornmayer},
\author[3]{D.~Kranich},
\author[1]{H.~Krawczynski},
\author[1]{H.~Lampeitl},
\author[7]{A.~Lindner},
\author[3]{E.~Lorenz},
\author[6]{N.~Magnussen},
\author[5]{O.~Mang},
\author[6]{H.~Meyer},
\author[3,4,2]{R.~Mirzoyan},
\author[4]{A.~Moralejo},
\author[4]{L.~Padilla},
\author[1]{M.~Panter},
\author[3]{R.~Plaga},
\author[7]{J.~Prahl},
\author[3]{C.~Prosch},
\author[5]{G.~Rauterberg},
\author[6]{W.~Rhode},
\author[7]{A.~R\"ohring},
\author[5]{M.~Samorski},
\author[4]{J.A.~Sanchez},
\author[5]{M.~Schilling},
\author[7]{D.~Schmele},
\author[6]{F.~Schr\"oder},
\author[5]{W.~Stamm},
\author[1]{H.J.~V\"olk},
\author[6]{B.~Wiebel-Sooth},
\author[1]{C.A.~Wiedner},
\author[5]{M.~Willmer}
 
\collab{HEGRA Collaboration}

\address[1]{Max-Planck-Institut f\"ur Kernphysik, P.O. Box 103980,
        D-69029 Heidelberg, Germany}
\address[2]{Yerevan Physics Institute, Yerevan, Armenia}
\address[3]{Max-Planck-Institut f\"ur Physik, F\"ohringer Ring 6,
        D-80805 M\"unchen, Germany}
\address[4]{Facultad de Ciencias Fisicas, Universidad Complutense,
         E-28040 Madrid, Spain}
\address[5]{Universit\"at Kiel, Inst. f\"ur Experim. u Angew. Physik,
       Olshausenstr.40, D-24118 Kiel, Germany}
\address[6]{BUGH Wuppertal, Fachbereich Physik, Gau\ss str.20,
        D-42119 Wuppertal, Germany}
\address[7]{Universit\"at Hamburg, II. Inst. f\"ur Experimentalphysik,
       Luruper Chaussee 149, D-22761 Hamburg, Germany}
\address[9]{Now at Forschungszentrum Karlsruhe, P.O. Box 3640, 76021 Karlsruhe}
\address{$^*$Corresponding author (Werner.Hofmann@mpi-hd.mpg.de\\
Fax +49 6221 516 603, Phone +49 6221 516 330)}

\maketitle
\clearpage

\begin{abstract}
For stereoscopic systems of imaging
atmospheric Cherenkov telescopes (IACTs), a key parameter
to optimize the sensitivity for VHE $\gamma$-ray point sources
is the intertelescope spacing.
Using pairs of telescopes of the HEGRA IACT system, the sensitivity
 of 
two-telescope stereo IACT systems is studied as a function
of the telescope spacing, ranging from 70~m to 140~m.
Data taken during the 1997 outburst of Mrk 501 are used
to evaluate both the detection rates before cuts, and the
sensitivity for weak signals after cuts to optimize the
sig\-nif\-i\-cance of signals. 
Detection rates decrease by about 1/3 between the minimum and
maximum spacing. The significance of signals is essentially
independent of distance, in the range investigated. 
\end{abstract}

\end{frontmatter}

\section{Introduction}

IACT stereoscopy --  the simultaneous observation of air showers with 
multiple imaging atmospheric Cherenkov telescopes (IACTs) under
different viewing angles -- has 
become the technique of choice for most of the next-generation
instruments for earth-bound gamma-ray astronomy in the VHE energy
range, such as VERITAS\cite{veritas} or HESS\cite{hess}. 
The stereoscopic observation of air showers allows improved 
determination of the direction of the primary and of its
energy, as well as a better suppression of backgrounds, compared
to single IACTs. A crucial question in the layout of stereoscopic
systems of IACTs is the spacing of the identical telescopes. Obviously, the
spacing of telescopes should be such that at least two telescopes
fit into the Cherenkov light pool with its typical diameter 
of around 250~m. For higher energy showers the light pool 
can increase considerably, but at large distances ($>$ 130 m) one
observes primarily light from shower particles scattered at larger
angles.
The coincidence rate between two telescopes
will decrease with increasing spacing; on the other hand, the
angle between the views and hence the quality of the stereoscopic
reconstruction of the shower geometry will improve. The opinions
within the IACT community
concerning the optimum geometry of IACT arrays differ; the HESS
project, for example, initially aimed for a spacing of 100~m between
adjacent telescopes; more recently, larger distances around
120~m or more are favored \cite{hess1,workshop1}. For the
otherwise similar VERITAS array, intertelescope distances
of initially 50~m, and later 80~m to 85~m were foreseen. Among existing
systems, the HEGRA 5-telescope IACT system has a characteristic spacing
of 70~m between its central telescope and the corner telescopes.
The WHIPPLE-GRANITE two-telescope system \cite{granite}
had a spacing of 140~m. The Telescope Array \cite{telescope_array}
uses a spacing of 120~m between the first three telescopes, and
later 70~m for the full array.

Optimization of IACT system geometry is heavily based on 
Monte-Carlo simulations. Over the last years, the quality
of these simulations has improved significantly, both
concerning the reliability of the shower simulation
\cite{sasha_kruger} and concerning the details of the
simulation of the telescopes and their readout (see, e.g,
\cite{hegra_mc}). Simulations have been tested extensively,
and key characteristics, such as the radial distribution
of Cherenkov light in the light pool, have been verified
experimentally \cite{hegra_pool}. Nevertheless, a
direct experimental verification of the dependence of the 
performance of IACT systems on the intertelescope distance would be highly
desirable.

The HEGRA IACT system at the Observatorio del Roque de los
Muchachos on La Palma consists of five telescopes, 
four of them arranged roughly in the form of a square with 100~m side
length, with the fifth telescope in the center. Selecting
pairs of telescopes, the  performance of two-telescope
stereo systems can be studied for distances between 70~m
(from the central telescope to the corner telescopes) and 
140~m (across the diagonal of the square), covering 
essentially the entire range of interest. This paper
reports the results of such a study, based on the large
sample of gamma-rays \cite{p501} acquired during the 1997 outburst of
Mrk 501.

\section{Telescope hardware and data set}

The HEGRA IACT system consists of five telescopes, each with
a tessellated mirror of 8.5~m$^2$ area and 5~m focal length,
and a 271-pixel camera with a 4.3$^\circ$ field of view.
The trigger condition requires a coincidence of two neighboring pixels
above a threshold $q_o$ to trigger the camera, and a 
coincidence of triggers from two cameras to record the data.
Details of the trigger system are given in \cite{lampeitl}.
Events are reconstructed by parameterizing the images
using the Hillas image parameters, and by geometrically
determining the direction and the impact point of the
shower. A simple and relatively efficient method for cosmic-ray
rejection is based on the  difference in the {\em width} of 
$\gamma$-ray and cosmic-ray images, respectively.
{\em Width} values are normalized to the mean {\em width} expected for
a gamma-ray image of a given intensity and impact distance,
and a {\em mean scaled width} is calculated by averaging over
telescopes. The $\gamma$-ray showers exhibit, by definition,
a {\em mean scaled width} around 1, whereas the broader cosmic-ray showers
show larger values. 
A cut requiring a {\em mean scaled width} below
1.2 or 1.3 keeps virtually all gamma-rays and rejects a 
significant fraction of cosmic rays; a cut at 1.0 to 1.1
has lower gamma-ray acceptance, but optimizes the significance
for the detection of $\gamma$-ray sources. 
The data analysis and performance of the
system are described in more detail in \cite{performance,p501}.

During 1997, when the data for this study were taken,
only four of the five telescopes were included in the HEGRA
IACT system; the fifth telescope (one of the corner telescopes)
was still equipped with an
older camera and was operated in stand-alone mode. The four
telescopes can be used to emulate six different
two-telescope stereo systems: three combinations with intertelescope
distances around 70~m (from the central telescope to the
three corner telescopes), two combinations with about 90~m
and 110~m (the sides of the imperfect square), and one
combination with 140~m (the diagonal). Data from pairs of
(triggered) telescopes are analyzed, ignoring the information
provided by the other telescopes. Since only two telescopes
are required to trigger the system, there is no trigger bias
or other influence from those other telescopes.

A slight difficulty arises since one compares combinations
of different telescopes, rather than varying the distance
between two given telescopes. While the four HEGRA system
telescopes used here are identical in their construction,
they differ somewhat in their age and hence the degree of mirror
deterioration, in the quality of the alignment of the mirror
tiles, and in the properties of the PMTs in the cameras,
which show systematic variations between production batches.
Different mirror reflectivities and PMT quantum efficiencies
result in slightly different energy thresholds for given (identical)
settings of electronics thresholds. The determination of
image {\em widths} and hence the background rejection are 
sensitive to the quality of the mirror adjustment. These
effects have to be determined from the data, and compensated.

The analysis is based on the data set recorded in 1997 during
the outburst of Mrk 501. Only data during the high-flux 
periods between MJD 50595 and MJD 50613 were used. The
data set was further restricted to small zenith angles, less 
than $20^\circ$, to approximate the situation for vertical
showers. Mrk 501
was observed in the so-called wobble mode, with the source
images 0.5$^\circ$ away from the center of the cameras.
An equivalent region imaged on the opposite side of the
camera center was used as off-source region. Given the angular
resolution of about 0.1$^\circ$, the on-source and off-source
regions are well separated.

\section{Analysis and results}

As a first step in the analysis, the detection rates were studied
as a function of intertelescope distance.

To equalize the energy thresholds of all telescopes,
pairs of telescopes were used to reconstruct showers and
events with cores located at equal distance from both
telescopes were selected. By comparing the mean {\em size}
of the images in the two telescopes, and in particular by
comparing the mean signal amplitude in the second-highest
pixel (which determines if a telescopes triggers or not),
and can derive correction factors which can be used to
equalize the response of the telescopes. Three of the
four telescopes were found to be identical within a few \%,
one had a sensitivity which was lower by about 25\%. In 
the analysis, pixel amplitudes are corrected correspondingly,
and only camera images with two pixels above 20
photoelectrons are accepted. This software threshold is high enough 
to eliminate any influence of the hardware threshold (at about
8 to 10 photoelectrons, depending on the data set), even after
the worst-case 25\% correction is applied.

\begin{figure}
\begin{center}
\mbox{
\epsfxsize10.0cm
\epsffile{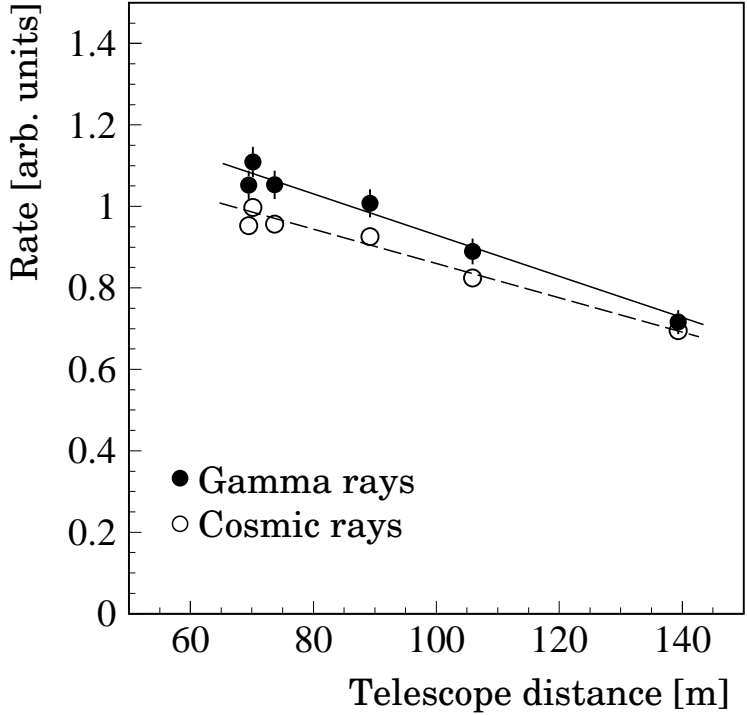}}
\caption{Detection rates of cosmic rays and of gamma rays as 
a function of the spacing between stereoscopic pairs of
telescopes, using a software trigger requirement
of more than 20 photoelectrons in two pixels of both
cameras. The normalization is arbitrary. Gamma-ray events
were selected using very loose cuts on pointing relative to
Mrk 501
($< 0.45^\circ$) and on the {\em mean scaled width} of the images
($< 1.3$), keeping essentially all gamma-rays. Cosmic-ray 
background in the on-source sample was subtracted on a 
statistical basis. The lines are drawn to guide the eye.}
\label{fig_rates}
\end{center}
\end{figure}

The resulting detection rates of cosmic rays and of gamma-rays 
-- after subtraction of the cosmic-ray background  -- are
shown in Fig.~\ref{fig_rates}. The rates measured for the
three different telescope combinations around 70~m spacing
agree within 5\%, indicating the precision in the adjustment
of telescope thresholds. Detection rates decrease with increasing
distance; between 70~m and 140~m, rates drop by about 1/3, 
with a very similar dependence for gamma rays as 
compared to cosmic rays. At first glance, this seems surprising,
since compared to gamma-ray showers,
 the distribution of Cherenkov light in proton-induced showers
-- and hence the probability to trigger -- is more strongly peaked
near the shower axis, favoring a smaller separation of telescopes.
On the other hand, however, the trigger probability for gamma-ray
showers near threshold - where most of the detection rate originates --
cuts off sharply around 120~m to 130~m, whereas proton-induced showers
occasionally trigger at larger distances \cite{koehler}. The two
effects compensate each other to a certain extent. In addition, 
about 33\% of the cosmic-ray triggers are caused by primaries heavier
than protons \cite{heavy}. These particles interact higher in the
atmosphere and produce a wider light pool, again enhancing the rate
for telescopes with wide spacing. In the later analysis, shape cuts
completely eliminate such events. Indeed, if only cosmic-ray events
with narrow (proton- or gamma-like) images are accepted, their rates
fall off steeper with distance than the all-inclusive rate, or the
gamma-ray rate.

\begin{figure}
\begin{center}
\mbox{
\epsfxsize10.0cm
\epsffile{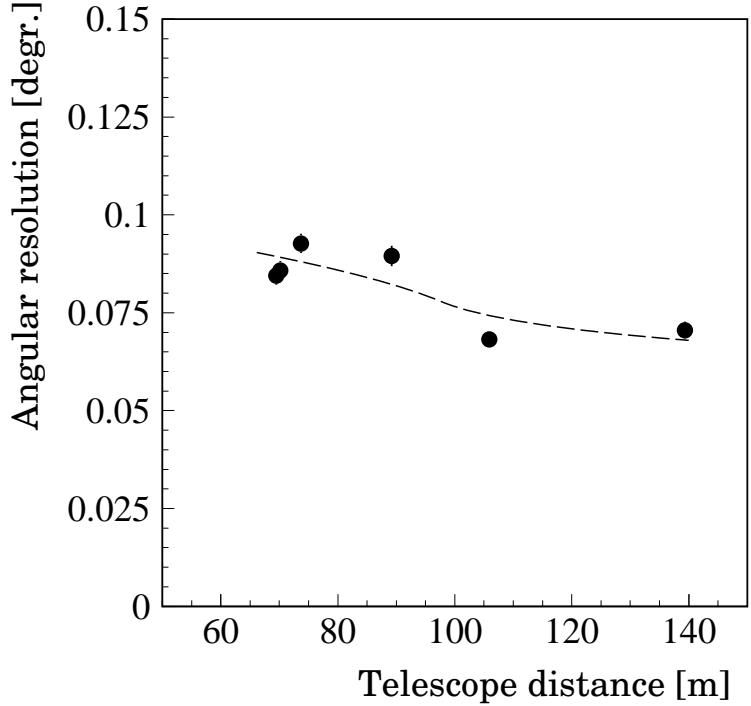}}
\caption{Angular resolution provided by pairs of telescopes
as a function of telescope spacing. A software trigger
threshold of 20 photoelectrons is used. The angular resolution
is determined by fitting a Gaussian to the difference between
shower direction and direction to the source, projected on two
orthogonal axes.
The dashed line 
is drawn to guide the eye.}
\label{fig_angres}
\end{center}
\end{figure}

As a measure of the sensitivity for weak sources, the ratio
$S/\sqrt{B}$ of the gamma-ray rate to the square root of 
the cosmic-ray rate was used. The significance of 
background-dominated signals scales with this ratio. For
optimum sensitivity, $S/\sqrt{B}$ is optimized by tighter
cuts on the pointing of showers, and on the image shapes.
Fig.~\ref{fig_angres} shows the angular resolution provided
by pairs of telescopes as a function of spacing. The three
pairs at 70~m show differences at the level of 10\%,
indicating small differences in the quality of the mirror alignment
and of the telescope alignment. Angular resolution
improves slightly with increasing spacing; however, given
the 10\% systematic variations, this effect is of marginal
significance.

\begin{figure}
\begin{center}
\mbox{
\epsfxsize10.0cm
\epsffile{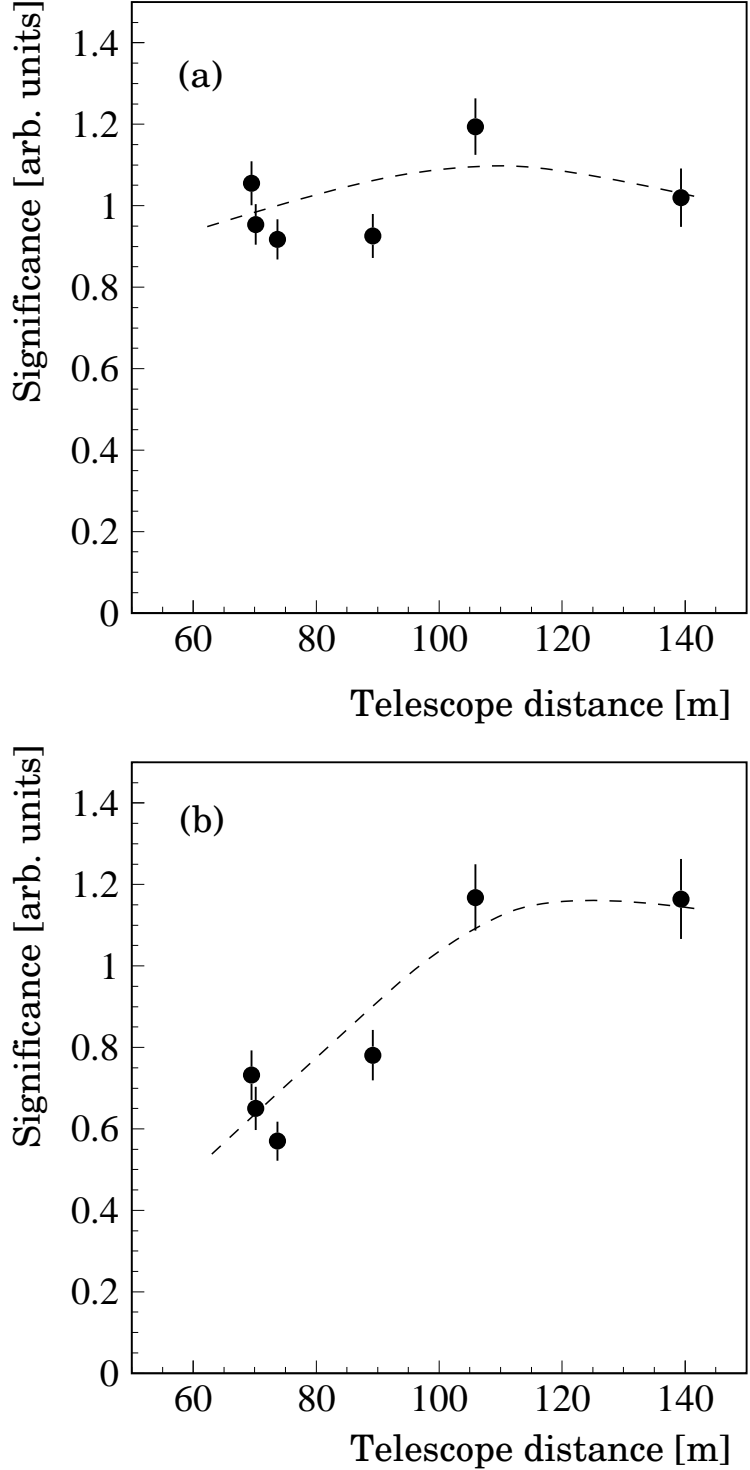}}
\caption{(a) Significance $S/\sqrt{B}$ of the Mrk 501 signal
for different pairs of telescopes, with optimized
cuts on shower pointing and image shapes in the two
telescopes. The significance is
normalized to $\approx 1$ for the 70~m spacing. The dashed line 
is drawn to guide the eye.
(b) As (a), but with an additional cut of a stereo angle
of at least $45^\circ$ between the two views. The normalization
if the same as for (a), so that the two data sets can be compared
directly. A weaker $20^\circ$ cut on the stereo angle, as
 used in some HEGRA analyses, has little effect on the
significance, and yields the almost same results as shown in (a).}
\label{fig_sign}
\end{center}
\end{figure}

To determine the sensitivity for the different combinations of
telescopes, cuts on pointing and on the {\em mean scaled
width} were optimized for each combination, resulting in pointing
cuts around $0.1^\circ$ and cuts on the {\em mean scaled width}
around 1.05. In addition, a lower limit of 0.75 was imposed
on the {\em mean scaled width}.
Due to differences in
the quality of the mirror alignment, the enhancement of 
significance due to such cuts differs slightly
between telescopes. This can be
studied by selecting events with telescopes $A$ and $B$,
and deriving the significance of the signal by cutting only
on the event shape in $A$, or only in $B$. For identical
telescopes, the results should be identical. Based on such
studies, sensitivity corrections for the different telescope
pairs were derived; these correction factors were always below
10\%. Fig.~\ref{fig_sign}(a) shows the resulting significance
(defined as $S/\sqrt{B}$) of the Mrk 501 signal. The errors
shown are statistical, and are dominated by the low statistics
in the cosmic-ray background sample after cuts (even though 
the background region was chosen to be larger than the signal region). 
One finds that the
significance is almost independent of telescope spacing, over the
70~m to 140~m range covered. 

Concerning systematic uncertainties, we note that the results
for the three combinations at 70~m agree reasonably well. Variations
of parameters such as the software trigger threshold, or the
exact values of the pointing cuts and angular cuts produce
stable results, with systematic variations of at most 10\%.

The absolute significance of a point source detected with
two stereoscopic IACTs, and its dependence on the intertelescope
distance will, of course, depend on additional cuts which may
be applied in the analysis. Additional requirements may, for example,
concern the stereo angle, defined as
the angle between the two views of the shower axis provided
by the two telescopes, or, equivalently, the angle between
the image axes in the two cameras.
If the stereo angle is small -- as it is
the case for events with shower impact points on or near the
line connecting the two telescopes, and for very distant showers
 -- the two views 
coincide and the spatial reconstruction of the shower axis
is difficult.
In some analyses, one will therefore add the requirement of a minimal
stereo angle between the two views, in order to increase
the reliability of the stereoscopic reconstruction. 
For a minimum stereo angle of $20^\circ$ -- as used in some HEGRA work --
the resulting significance remains virtually unchanged compared
to the data shown in Fig.~\ref{fig_sign}(a).
With a large minimum stereo angle of $45^\circ$
(Fig.~\ref{fig_sign}(b)), the distance dependence becomes more
pronounced; for small telescope distances, significance is reduced,
while for large distances it remains or is even slightly enhanced.
The
explanation is simple. For a telescope spacing $d$ and
a stereo angle greater than $45^\circ$, shower cores
are accepted at most up to a distance of $\approx 1.2 d$ from the
center of the telescope pair. As long as this distance is 
 below the radius of the light pool, the effective detection area
 is reduced. 
One concludes that such a cut might be beneficial to reduce 
systematic effects in spectral measurements, for example, but should not be
applied in searches for gamma-ray sources.

Another quantity which influences the distance dependence is the field
of view of the cameras. Cameras with small field of view will perform
worse for large distances. For example, rejecting events with image
centroids beyond $1^\circ$ from the camera center reduces the 
sensitivity for the $140$ m point by a factor of about 2.

\section{Summary}

The performance of two-telescope stereoscopic IACT systems 
was studied experimentally using the HEGRA telescopes, as a function of
telescope spacing in the range between 70~m and 140~m.
While detection rates decrease with increasing spacing, the
significance of source signals is almost independent of
distance, with a slight improvement for distances  beyond 100~m.
These results confirm Monte Carlo simulations (see, for example, 
\cite{workshop1})
which generally show that the exact choice of telescope spacing
is not a very critical parameter in the design of IACT systems
(provided that the field of view of the cameras is sufficiently
large not to limit the range of shower impact parameters).
An analysis requirement of a minimum angle between views favors
larger distances.

The results shown apply, strictly speaking, only to two-telescope
systems. In IACT arrays with a large number of telescopes,
one may place telescopes at maximum
spacing in order to maximize the effective area of the array,
under the condition that most individual showers are observed by
two or maybe three or four telescopes. Alternatively one
may choose to place telescopes close to each other, such that 
an individual shower is observed simultaneously by almost all telescopes,
improving the quality of the reconstruction and lowering the
threshold, at the expense of detection area at energies well
above threshold. In the first case, one would probably use a
spacing between adjacent telescopes between 100~m and 150~m;
in the second case, one would limit the maximum spacing 
between any pair of telescopes to this distance.

\section*{Acknowledgements}

The support of the HEGRA experiment by the German Ministry for Research 
and Technology BMBF and by the Spanish Research Council
CYCIT is acknowledged. We are grateful to the Instituto
de Astrofisica de Canarias for the use of the site and
for providing excellent working conditions. We gratefully
acknowledge the technical support staff of Heidelberg,
Kiel, Munich, and Yerevan.

\end{document}